\documentclass{article}
\usepackage[T1]{fontenc}
\usepackage[utf8]{inputenc}
\usepackage{ismir}
\usepackage{amsmath,cite,url}
\usepackage{graphicx}
\usepackage{color}

\usepackage{amssymb}
\usepackage{multirow}
\usepackage{booktabs}
\usepackage[ruled,vlined,linesnumbered]{algorithm2e}
\usepackage{enumitem}
\usepackage{booktabs} 
\usepackage{makecell}  

\title{Song Aesthetics Evaluation with Multi-Stem Attention and Hierarchical Uncertainty Modeling}

\multauthor
  {Yishan Lv$^1$ \hspace{0.45cm} Jing Luo$^1$ \hspace{0.45cm} Boyuan Ju$^1$}
  {{\bf Yang Zhang$^2$ \hspace{0.45cm} Xinda Wu$^2$ \hspace{0.45cm} Bo Yuan$^2$ \hspace{0.45cm} Xinyu Yang$^{1,*}$}\\
   $^1$ School of Computer Science and Technology, Xi'an Jiaotong University, Xi'an, China\\
   $^2$ Central Media Technology Institute, Huawei\\
   {\tt\small yisan@stu.xjtu.edu.cn, yxyphd@mail.xjtu.edu.cn}}

\def\authorname{Y. Lv, J. Luo, B. Ju, Y. Zhang, X. Wu, B. Yuan, and X. Yang}

\begin{document}

\maketitle
\renewcommand{\thefootnote}{*}
\footnotetext{Corresponding author}
\renewcommand{\thefootnote}{\arabic{footnote}}

\begin{abstract}
Music generative artificial intelligence (AI) is rapidly expanding music content, necessitating automated song aesthetics evaluation for efficient music retrieval and recommendation. However, existing studies largely focus on speech, audio or singing quality, leaving song aesthetics underexplored. Moreover, conventional approaches often predict a precise Mean Opinion Score (MOS) value directly, which struggles to capture the nuances of human perception in song aesthetics evaluation. This paper proposes a song-oriented aesthetics evaluation framework, featuring two novel modules: 1) Multi-Stem Attention Fusion (MSAF) builds bidirectional cross-attention between mixture-vocal and mixture-accompaniment pairs, fusing them to capture complex musical features; 2) Hierarchical Granularity-Aware Interval Aggregation (HiGIA) learns multiple granularity score probability distributions, aggregates them into a score interval, and applies a regression within the interval to produce the final score. We evaluated on two datasets of full-length songs: SongEval dataset (AI-generated) and an internal aesthetics dataset (human-created), and compared with two state-of-the-art (SOTA) models. Results show that the proposed method achieves stronger performance for multi-dimensional song aesthetics evaluation.
The inference code and checkpoint are publicly available at \url{https://github.com/yisan33/song-aesthetics-evaluation}.
\end{abstract}

\section{Introduction}\label{sec:introduction}

The advent of music generative AI\cite{copet2023simple, chen2025diffrhythm+, yuan2026yue, lei2024songcreator} is rapidly increasing the volume of music releases, making manual quality screening and manual aesthetics evaluation impractical. This highlights the urgency for automated song aesthetics evaluation to enable efficient music retrieval and recommendation. Such systems could provide valuable feedback to guide music creation and improve overall quality. 

Deep learning has been actively explored for image quality \cite{schultze2023explaining, pang2025blockiqa}, for assessing specific music attributes such as timbre quality \cite{acquilino2023dataset} and memorability \cite{tseng2024dataset}, and for audio quality assessment, particularly in speech and singing. In speech quality assessment, NISQA \cite{mittag2021nisqa} introduces self-attention into a CNN-based architecture for multi-dimensional assessment;
LDNet \cite{huang2022ldnet} models listener dependence to stabilize predictions;
and UTMOS \cite{saeki2022utmos} uses self-supervised learning with model ensembling to enhance stability.
In singing quality assessment, 
Zhang et al. \cite{zhang2021learn} introduce reference-based deep metric learning for multi-dimensional singing evaluation; 
PS-SQA \cite{shi2024pitch} employs pitch and spectrum-aware predictors to enhance performance;
and Ju et al. \cite{ju2024end} use cross-attention with data augmentation for singing skill evaluation.
In music-related evaluation,
MusicEval \cite{liu2025musiceval} provides a benchmark for the evaluation of short, accompaniment-only music clips and serves as the dataset for Track 1 of the AudioMOS Challenge 2025 \cite{huang2025audiomos};
MMMOS \cite{lin2025mmmos} evaluates clip-level text-to-speech, audio, and music;
Audiobox \cite{tjandra2025meta} benchmarks multi-domain audio aesthetics;
and SongEval \cite{yao2025songeval} provides a benchmark dataset with baselines for song aesthetics.

However, these studies mostly do not target full-length song aesthetics evaluation. Simply transferring speech or singing frameworks to the task of song aesthetics evaluation yields limited effectiveness for two main reasons:
1) Compared to speech, songs present a more complex structure, consisting of interwoven vocal and accompaniment stems. Compared to singing, a song's aesthetic evaluation also depends on arrangement, melodic structure and production, instead of just the vocal performance \cite{moore2016song};
2) Traditional approaches predict a precise MOS directly. In contrast, human experts follow a coarse-to-fine process: they first estimate an approximate score interval under subjective uncertainty \cite{mcpherson2004measuring} and only then decide a precise score. This mismatch contradicts the human strategy, compromising stability and accuracy.

To overcome these two challenges of musical complexity and evaluation uncertainty, this paper proposes a full-length song aesthetics evaluation framework, featuring two novel modules: 
1) Multi-Stem Attention Fusion (MSAF) utilizes bidirectional cross-attention between the mixture and the vocal and accompaniment stems, and then fuses the three representations to model the complex interplay across stems and to capture richer musical features.
2) Hierarchical Granularity-Aware Interval Aggregation (HiGIA) trains three hierarchical classifiers to obtain multi-granularity score distributions, aggregates high-confidence bins into a consensus interval, and then performs regression within the interval to produce the final score. This coarse-to-fine process achieves more stable and accurate predictions.
To evaluate the proposed model, we conducted experiments on two datasets of full-length songs: 1) the SongEval dataset (mostly AI-generated) and 2) an internal aesthetics dataset (mostly human-created). Our framework outperforms two strong baselines across all four standard metrics. We also conducted ablation studies to analyze the effectiveness of MSAF and HiGIA.

The main contributions of this work are summarized as follows:
\begin{itemize} 
    \item We propose a novel framework specifically designed for full-length song aesthetics evaluation. The framework consists of two modules, MSAF and HiGIA, to address the challenges of musical complexity and subjective evaluation uncertainty.
    \item We design the Multi-Stem Attention Fusion (MSAF) module, which utilizes bidirectional cross-attention to explicitly model the interplay among the mixture, vocal, and accompaniment stems, capturing complementary musical cues.
    \item We introduce the Hierarchical Granularity-Aware Interval Aggregation (HiGIA) module, which aggregates multi-granularity score distributions into a consensus interval for final score regression. This coarse-to-fine process improves prediction stability and accuracy.

\end{itemize}

\section{Method}

\subsection{Model Overview}

We first apply music source separation to the input full-length song (i.e., a complete audio recording that contains both vocals and accompaniment, typically lasting 3 to 5 minutes) to obtain the vocal and accompaniment stems. Together with the original mixture signal, these components form three parallel inputs, namely the mixture, vocal, and accompaniment branches. 

Each branch is fed into the pretrained MuQ \cite{zhu2025muq} encoder to extract musical representations. We employ a CBAM-based \cite{woo2018cbam} attention-weighted sum over the multi-layer hidden outputs of MuQ to obtain a single embedding per stem. These embeddings are then passed through Transformer encoders to further capture temporal dependencies and contextual information within each branch, yielding stem-specific representations for subsequent fusion. 

\begin{figure}[!tbp]
    \centering 
    \includegraphics[width=0.91\linewidth]{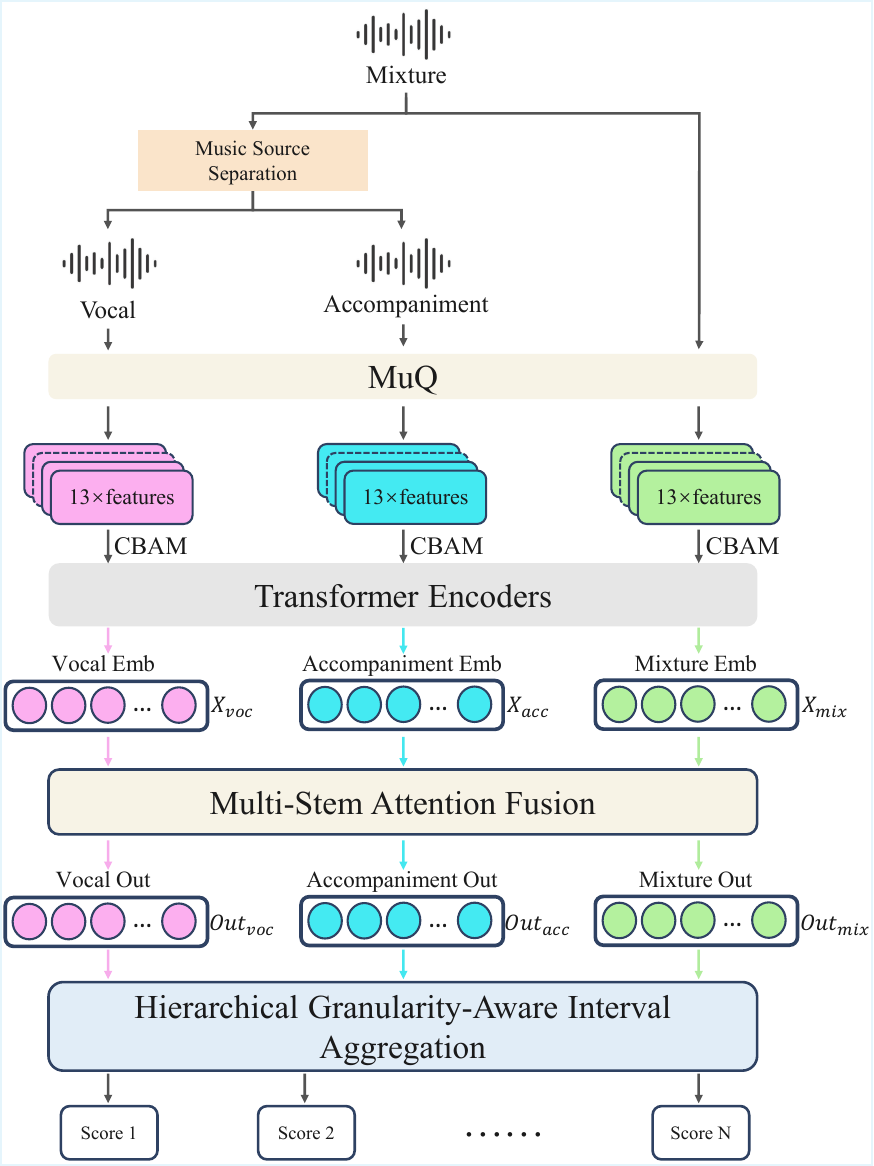}
    \caption{Overall architecture of the proposed framework. The input song is separated into mixture, vocal, and accompaniment stems, which are encoded into stem-specific representations. These features are then processed by MSAF and HiGIA to predict final aesthetic scores.}
    \label{fig:architecture}
\end{figure}

The resulting features are then fused by the Multi-Stem Attention Fusion, which explicitly models the interactions among the mixture, vocal, and accompaniment representations to capture complementary musical cues. Finally, the Hierarchical Granularity-Aware Interval Aggregation processes this fused representation by first producing multi-granularity score distributions and then aggregating them for final regression, yielding the multi-dimensional aesthetic score predictions. Figure~\ref{fig:architecture} illustrates the model architecture.

\subsection{Multi-Stem Attention Fusion}

The feature representations for the mixture, vocal stem, and accompaniment stem, output by Transformer encoders, are denoted as $X_{\text{mix}}, X_{\text{voc}}, X_{\text{acc}} \in \mathbb{R}^{T \times d}$, where $T$ is the number of frames, and $d$ is the feature dimension.

To model interactions between stems, we adopt a bidirectional cross-attention mechanism \cite{hiller2024perceiving}. Instead of constructing queries, keys, and values for all stems, we designate the mixture stem as the central query source. Specifically, the mixture stem generates queries and values, while the vocal and accompaniment stems provide keys and values. The linear projections are formulated as:
\begin{equation}\label{eq:qkv_projections}
\begin{aligned}
    Q_{\text{mix}} &= X_{\text{mix}}W_{\text{mix}}^Q, & V_{\text{mix}} &= X_{\text{mix}}W_{\text{mix}}^V, \\
    K_{\text{voc}} &= X_{\text{voc}}W_{\text{voc}}^K, & V_{\text{voc}} &= X_{\text{voc}}W_{\text{voc}}^V, \\
    K_{\text{acc}} &= X_{\text{acc}}W_{\text{acc}}^K, & V_{\text{acc}} &= X_{\text{acc}}W_{\text{acc}}^V,
\end{aligned}
\end{equation}
where $W_{(\cdot)}^Q, W_{(\cdot)}^K, W_{(\cdot)}^V \in \mathbb{R}^{d \times d_k}$ are the branch-specific learnable weight matrices, and $d_k$ is the projection dimension. 

Based on the scaled dot-product attention \cite{vaswani2017attention}, we compute a shared similarity matrix between the mixture and each of the vocal and accompaniment stems: 
\begin{equation}\label{eq:attention_scores}
\begin{aligned}
    S_{\text{mv}} &= \frac{Q_{\text{mix}}K_{\text{voc}}^{\top}}{\sqrt{d_k}} = S_{\text{vm}}^{\top}, \\
    S_{\text{ma}} &= \frac{Q_{\text{mix}}K_{\text{acc}}^{\top}}{\sqrt{d_k}} = S_{\text{am}}^{\top}.
\end{aligned}
\end{equation}

From the shared similarity matrices, we compute the bidirectional cross-attention outputs between the mixture and each stem, $\operatorname{Attn}_{\text{mix}}^{\text{s}}$ and $\operatorname{Attn}_{\text{s}}^{\text{mix}}$, where $s \in \{\text{voc}, \text{acc}\}$. By transposing the shared similarity matrices (i.e., $S_{\text{vm}} = S_{\text{mv}}^{\top}$ and $S_{\text{am}} = S_{\text{ma}}^{\top}$), bidirectional interactions can be modeled without introducing additional query projections. The outputs are computed as follows:
\begin{equation}\label{eq:bidirectional_attention}
\begin{aligned}
    \operatorname{Attn}_{\text{mix}}^{\text{voc}} &= \operatorname{softmax}(S_{\text{mv}})\, V_{\text{voc}}, \\
    \operatorname{Attn}_{\text{voc}}^{\text{mix}} &= \operatorname{softmax}(S_{\text{vm}})\, V_{\text{mix}}, \\
    \operatorname{Attn}_{\text{mix}}^{\text{acc}} &= \operatorname{softmax}(S_{\text{ma}})\, V_{\text{acc}}, \\
    \operatorname{Attn}_{\text{acc}}^{\text{mix}} &= \operatorname{softmax}(S_{\text{am}})\, V_{\text{mix}}.
\end{aligned}
\end{equation}

\begin{figure}[!tbp]
    \centering
    \includegraphics[width=1\linewidth]{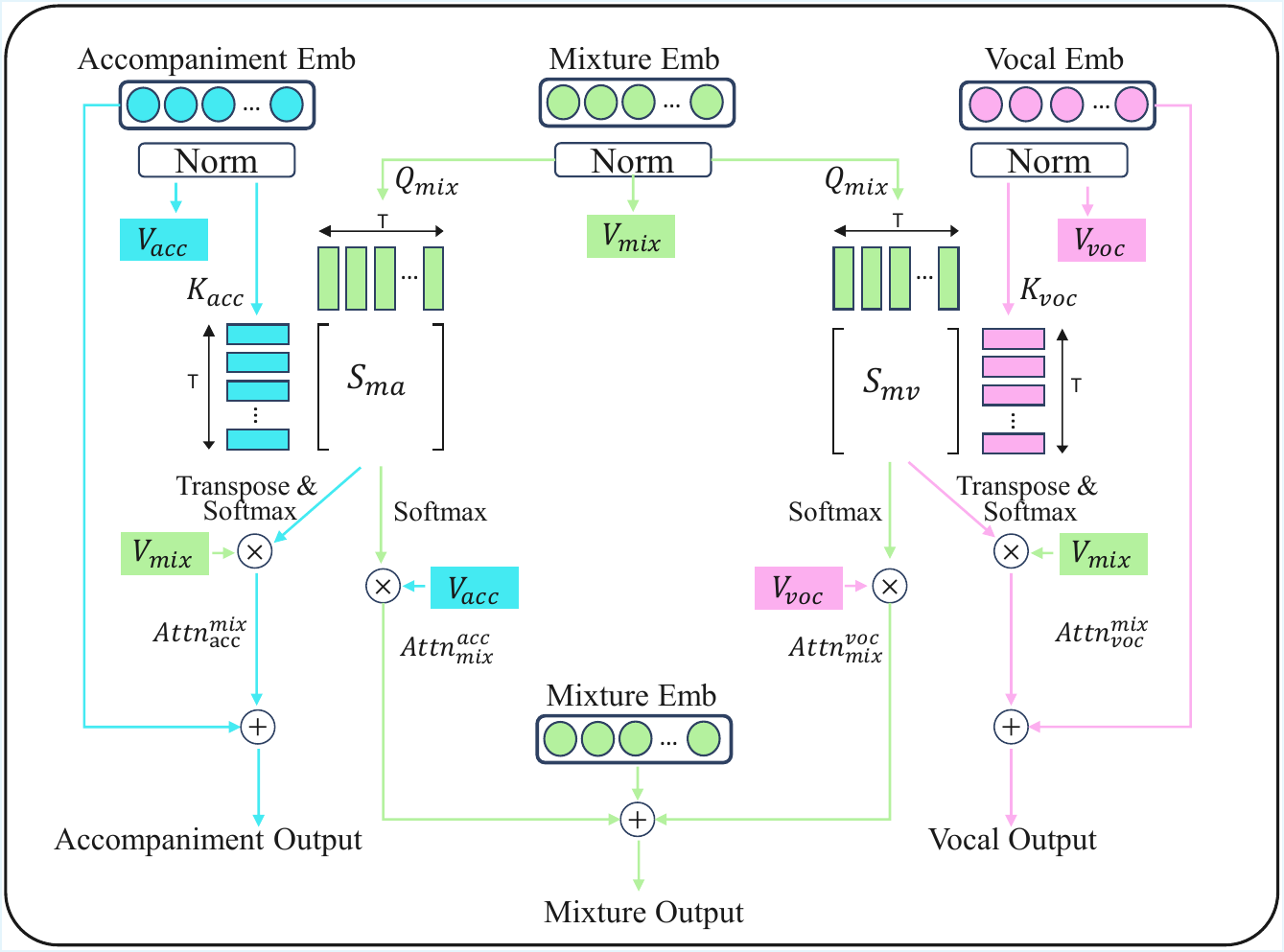}
    \caption{Illustration of the MSAF module. The mixture representation interacts bidirectionally with both the vocal and accompaniment branches, enabling the framework to capture complementary cross-stem musical cues.}
    \label{fig:multi-stem}
\end{figure}

Finally, residual connections yield the outputs for the MSAF. Figure~\ref{fig:multi-stem} illustrates the process.
\begin{equation}\label{eq:update_mix}
\begin{gathered}
  \text{Out}_{\text{mix}} = X_{\text{mix}} + \operatorname{Attn}_{\text{mix}}^{\text{voc}} + \operatorname{Attn}_{\text{mix}}^{\text{acc}}, \\
  \text{Out}_{\text{voc}} = X_{\text{voc}} + \operatorname{Attn}_{\text{voc}}^{\text{mix}}, \\
  \text{Out}_{\text{acc}} = X_{\text{acc}} + \operatorname{Attn}_{\text{acc}}^{\text{mix}}.
\end{gathered}
\end{equation}

\begin{algorithm}[!t]
\caption{Granularity-Aware Score Interval Generation}
\label{alg:interval_gen}
\KwIn{%
  Candidate class set of segment $i$: 
  $S_i = \{\,c_{g,k} \mid p_{g,k} > 1/K_g\}$ 
  \\
  Mapping from class to score range $r(c) = [l(c),\, u(c)]$
}
\KwOut{Adaptive score interval $RI_i = [L,\,U]$}
\BlankLine
\textbf{Initialise:} $O \leftarrow \varnothing$\; $I \leftarrow \varnothing$\;
\BlankLine
$\Phi \leftarrow \binom{|S_i|}{2}$ 
\ForEach{pair $(c_a,c_b)$ among the $\Phi$ combinations}{
    \lIf{$r(c_a)\cap r(c_b)\neq\varnothing$}{
        $O \leftarrow O \cup \{c_a,c_b\}$}
}
$I \leftarrow S_i \setminus O$\;
\BlankLine
\lIf{$|O| > |I|$}{            
  $L = \dfrac{1}{|O|}\,\sum_{c\in O} l(c)$\;
  $U = \dfrac{1}{|O|}\,\sum_{c\in O} u(c)$
}
\Else{                           
  $w_{O} = |O| / |S_i|$;\quad $w_{I}=1-w_{O}$\;
  $L = w_{O}\,\dfrac{1}{|O|}\sum_{c\in O} l(c) + w_{I}\,\min_{c\in I} l(c)$\;
  
  $U = w_{O}\,\dfrac{1}{|O|}\sum_{c\in O} u(c) + w_{I}\,\max_{c\in I} u(c)$\;
}
\Return{$RI_i = [L,\,U]$}
\end{algorithm}

\subsection{Hierarchical Granularity-Aware Interval Aggregation}

For each aesthetic dimension of a song, we set up three classifiers at different granularities, $\{C_{g_1},C_{g_2},C_{g_3}\}$, corresponding to coarse($g_1$), medium($g_2$), and fine($g_3$) levels. At a given granularity $g$, the classifier $C_g$ discretizes the continuous score into $K_g$ score bins, denoted as $c_{g,1},...,c_{g,K_g}$. Each score bin $c_{g,k}$ maps to a specific score interval $r(c_{g,k}) = [l(c_{g,k}), u(c_{g,k})]$. Finer granularities utilize more bins, i.e., $K_{g_1}<K_{g_2}<K_{g_3}$, thereby producing tighter intervals.
For instance, on a 100-point scale we can set $K_{g_1} = 3$ with 33-point bins $[0,33), [33,66),[66,100)$; $K_{g_2}=5$ with 20-point bins; $K_{g_3}=10$ with 10-point bins.

We apply softmax to the output logits $z^{(g)}$ of each classifier $C_g$ to obtain the posterior probability distribution, from which we then select a set $S$ of candidate score bins.
\begin{gather}
    \label{eq:posterior_properties}
    p^{(g)} = \operatorname{softmax}(z^{(g)}), \quad \sum_{k=1}^{K_g} p_k^{(g)} = 1 \\
    \label{eq:candidate_set}
    S = \{ c_{g,k} \mid p_k^{(g)}>1/K_g\}.
\end{gather}

We select the score bins that have overlapping intervals to form an overlap set $O$. The complement of this set is then taken to form the isolated set $I= S \setminus O$.
\begin{equation}
    \label{eq:set_partition_centered}
    O = \{ c \in S \mid \exists c^{\prime} \in S, c^{\prime} \neq c, r(c) \cap r(c^{\prime}) \neq \varnothing \}
\end{equation}

When the overlap set $O$ is larger than the isolated set $I$ (i.e., $|O|>|I|$), we obtain the score interval $[L, U]$ by aggregating the lower/upper bounds over $O$ \cite{dong2021hierarchical}.
\begin{equation}\label{eq:interval_definition}
\begin{gathered}
    L = \frac{1}{|O|} \sum_{c \in O} l(c), \\
    U = \frac{1}{|O|} \sum_{c \in O} u(c).
\end{gathered}
\end{equation}

Otherwise, (i.e., $|O| \leq |I|$), indicating more disagreement among the classifiers, we adopt a more conservative aggregation of $O$ and $I$ using weights $w_O = {|O|}/{|S|}$ and $w_I = 1 - w_O$. The complete process is detailed in Algorithm~\ref{alg:interval_gen}.
\begin{equation}\label{eq:helper_definitions}
\begin{aligned}
    L &= w_O\left(\frac{1}{|O|} \sum_{c \in O} l(c)\right) + w_I \min_{c \in I} l(c) \\
    U &= w_O\left(\frac{1}{|O|} \sum_{c \in O} u(c)\right) + w_I \max_{c \in I} u(c)
\end{aligned}
\end{equation}

\begin{figure}[!tbp]
    \centering
    \includegraphics[width=1\linewidth]{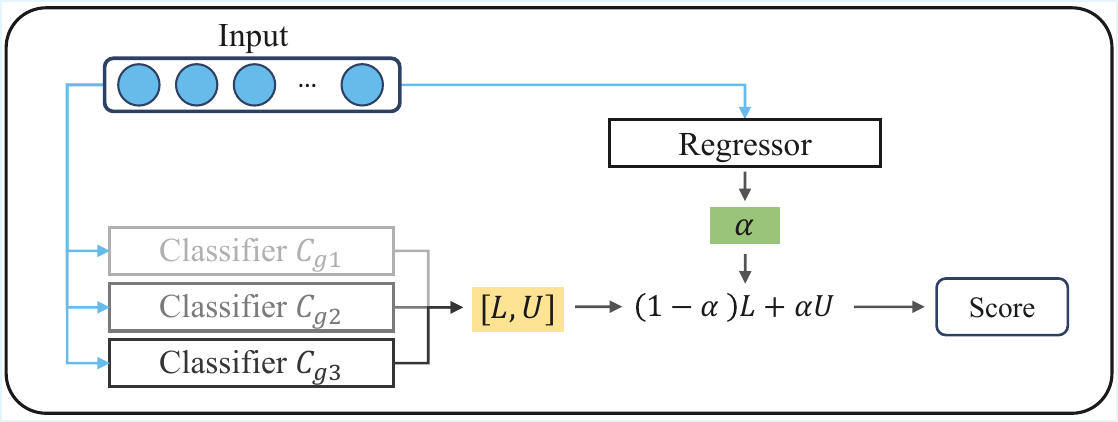}
    \caption{Illustration of the HiGIA module. Following a coarse-to-fine process, it first aggregates multi-granularity distributions into a consensus interval $[L, U]$, and then predicts the final aesthetic score within this interval.}
    \label{fig:Hierg}
\end{figure}

Given $[L, U]$, an MLP regressor is employed to predict an interpolation coefficient $\alpha \in [0,1]$, which determines the final aesthetic score. The entire coarse-to-fine process is illustrated in Figure~\ref{fig:Hierg}.
\begin{equation}
    \label{eq:interpolation}
    \hat{y} = (1 - \alpha) \cdot L + \alpha \cdot U
\end{equation}

\section{EXPERIMENT}

\subsection{Dataset}

To validate the effectiveness and generalization ability of our proposed framework, we conduct experiments on two distinct datasets covering both AI-generated and human-created music, as summarized in Table\ref{tab:dataset_comparison}. 

\begin{itemize}[leftmargin=*, itemsep=0pt, parsep=0pt, topsep=4pt]
    \item \textbf{SongEval aesthetics evaluation dataset} \cite{yao2025songeval}, an open-source collection containing 2,399 full-length songs in English and Chinese, the majority of which are AI-generated, each annotated on five dimensions: \textit{Musicality}, \textit{Coherence}, \textit{Memorability}, \textit{Clarity}, and \textit{Naturalness}, on a 5-point scale. This dataset provides a benchmark for assessing model performance on AI-generated music. 

    \item \textbf{Internal aesthetics evaluation dataset}, containing 1,569 full-length Chinese songs, most of which are human-created. The dataset is annotated by four music experts across five aesthetic dimensions, with each dimension scored on a 100-point scale: (1) \textit{Overall}, referring to the overall musical quality of the song; (2) \textit{Singing}, referring to timbre and vocal technique in conveying emotion; (3) \textit{Melody}, referring to the integrity, flow, and memorability of the melodic line; (4) \textit{Arrangement}, referring to instrumental coordination and textural layering; and (5) \textit{Audio Quality}, referring to clarity, vocal-accompaniment balance, and stereo imaging. This internal dataset enables a fine-grained evaluation of human-created songs, providing a complementary setting for examining the robustness of the proposed method.
\end{itemize}

\subsection{Baseline Models}

To validate the advantages of the proposed framework, we select two state-of-the-art (SOTA) models that demonstrated the strongest performance on the SongEval benchmark for comparison:

\begin{itemize}[leftmargin=*, itemsep=0pt, parsep=0pt, topsep=4pt]
    \item \textbf{SSL-based}: An SSL-based model following the design of \cite{cooper2022generalization}, adapted for song aesthetics by replacing the input features with those extracted from MuQ. 
    
    \item \textbf{UTMOS-based}: Based on the UTMOS framework \cite{saeki2022utmos}, an ensemble learning system that won the VoiceMOS 2022 Challenge \cite{huang2022voicemos}.
\end{itemize}

\begin{table}[t] 
\centering
\caption{Comparison of dataset characteristics between SongEval Benchmark and the Internal Dataset.}
\label{tab:dataset_comparison}
\renewcommand{\arraystretch}{1.2} 
\resizebox{\columnwidth}{!}{
\begin{tabular}{lcc} 
\hline
\textbf{Feature} & \textbf{SongEval Dataset} & \textbf{Internal Dataset} \\ \hline
Volume & 2,399 Full-length Songs & 1,569 Full-length Songs \\ \hline
Origin & Mostly AI-generated & Mostly Human-created \\ \hline
Language & English \& Chinese & Chinese \\ \hline
Annotators & 16 experts & 4 experts \\ \hline
\makecell[l]{Aesthetic \\ Dimensions} & 
\makecell{Musicality \\ Coherence \\ Memorability \\ Clarity \\ Naturalness} & 
\makecell{Overall \\ Singing \\ Melody \\ Arrangement \\ Audio Quality} \\ \hline
Scale & 5-point Scale & 100-point Scale \\ \hline
\end{tabular}
}
\end{table}

\begin{table*}[!t]
  \centering
  \caption{Comparison with baselines on SongEval and the Internal dataset. $\uparrow$/$\downarrow$ indicate higher/lower is better. \textbf{Bold} and \underline{underlined} denote the best and second-best results.}
  \label{tab:comparison-results}
  
  \renewcommand{\arraystretch}{0.88} 
  \small
  \setlength{\tabcolsep}{1.8pt} 
  
  \begin{tabular*}{\textwidth}{@{\extracolsep{\fill}} lccccc | lccccc }
    \toprule
    \multicolumn{6}{c}{\textbf{SongEval Dataset (5-point scale)}} & \multicolumn{6}{c}{\textbf{Internal Dataset (100-point scale)}} \\
    \cmidrule(r){1-6} \cmidrule(l){7-12}
    \textbf{Dim} & \textbf{Model} & \textbf{MSE$\downarrow$} & \textbf{LCC$\uparrow$} & \textbf{SRCC$\uparrow$} & \textbf{KTAU$\uparrow$} & 
    \textbf{Dim} & \textbf{Model} & \textbf{MSE$\downarrow$} & \textbf{LCC$\uparrow$} & \textbf{SRCC$\uparrow$} & \textbf{KTAU$\uparrow$} \\
    \midrule
    
    \multirow{3}{*}{Musicality}
      & SSL-based   & 0.277 & 0.895 & 0.891 & 0.721 & 
      \multirow{3}{*}{Overall}
      & SSL-based   & 27.0 & 0.889 & 0.888 & 0.701 \\
      & UTMOS-based & \underline{0.252} & \underline{0.899} & \underline{0.896} & \underline{0.725} & 
      & UTMOS-based & \underline{25.6} & \underline{0.892} & \underline{0.889} & \underline{0.702} \\
      & \textbf{Ours} & \textbf{0.239} & \textbf{0.902} & \textbf{0.906} & \textbf{0.744} & 
      & \textbf{Ours} & \textbf{21.7} & \textbf{0.912} & \textbf{0.921} & \textbf{0.753} \\
    \midrule
    
    \multirow{3}{*}{Coherence}
      & SSL-based   & 0.253 & 0.891 & 0.887 & 0.718 & 
      \multirow{3}{*}{Singing}
      & SSL-based   & 35.5 & 0.856 & 0.873 & 0.685 \\
      & UTMOS-based & \underline{0.250} & \underline{0.893} & \underline{0.888} & \underline{0.720} & 
      & UTMOS-based & \underline{32.0} & \underline{0.857} & \underline{0.876} & \underline{0.687} \\
      & \textbf{Ours} & \textbf{0.243} & \textbf{0.895} & \textbf{0.896} & \textbf{0.732} & 
      & \textbf{Ours} & \textbf{27.3} & \textbf{0.885} & \textbf{0.907} & \textbf{0.736} \\
    \midrule
    
    \multirow{3}{*}{Memorability}
      & SSL-based   & \underline{0.355} & 0.877 & 0.869 & \underline{0.694} & 
      \multirow{3}{*}{Melody}
      & SSL-based   & \underline{31.7} & 0.812 & 0.802 & 0.614 \\
      & UTMOS-based & \textbf{0.321} & \underline{0.881} & \textbf{0.875} & \textbf{0.702} & 
      & UTMOS-based & \underline{31.7} & \underline{0.815} & \underline{0.804} & \underline{0.618} \\
      & \textbf{Ours} & \textbf{0.321} & \textbf{0.882} & \underline{0.874} & \textbf{0.702} & 
      & \textbf{Ours} & \textbf{29.4} & \textbf{0.828} & \textbf{0.830} & \textbf{0.643} \\
    \midrule
    
    \multirow{3}{*}{Clarity}
      & SSL-based   & 0.291 & 0.884 & 0.879 & 0.699 & 
      \multirow{3}{*}{Arrangement}
      & SSL-based   & 28.0 & 0.839 & 0.820 & 0.636 \\
      & UTMOS-based & \underline{0.278} & \underline{0.887} & \underline{0.882} & \underline{0.702} & 
      & UTMOS-based & \textbf{16.0} & \textbf{0.861} & \textbf{0.826} & \underline{0.637} \\
      & \textbf{Ours} & \textbf{0.264} & \textbf{0.893} & \textbf{0.891} & \textbf{0.717} & 
      & \textbf{Ours} & \underline{18.9} & \underline{0.843} & \underline{0.824} & \textbf{0.650} \\
    \midrule
    
    \multirow{3}{*}{Naturalness}
      & SSL-based   & 0.285 & 0.885 & 0.879 & 0.702 & 
      \multirow{3}{*}{Audio Quality}
      & SSL-based   & 23.3 & 0.894 & \underline{0.886} & \underline{0.710} \\
      & UTMOS-based & \textbf{0.259} & \textbf{0.892} & \textbf{0.889} & \textbf{0.714} & 
      & UTMOS-based & \textbf{20.3} & \underline{0.897} & \underline{0.886} & \underline{0.710} \\
      & \textbf{Ours} & \underline{0.263} & \underline{0.889} & \underline{0.883} & \underline{0.709} & 
      & \textbf{Ours} & \underline{21.1} & \textbf{0.915} & \textbf{0.910} & \textbf{0.742} \\
    \midrule
    
    \multirow{3}{*}{Average}
      & SSL-based   & 0.292 & 0.886 & 0.881 & 0.707 & 
      \multirow{3}{*}{Average}
      & SSL-based   & 29.1 & 0.858 & 0.854 & 0.669 \\
      & UTMOS-based & \underline{0.273} & \underline{0.890} & \underline{0.886} & \underline{0.713} & 
      & UTMOS-based & \underline{25.1} & \underline{0.864} & \underline{0.856} & \underline{0.671} \\
      & \textbf{Ours} & \textbf{0.266} & \textbf{0.892} & \textbf{0.890} & \textbf{0.721} & 
      & \textbf{Ours} & \textbf{23.7} & \textbf{0.877} & \textbf{0.878} & \textbf{0.705} \\
    \bottomrule
  \end{tabular*}
  \vspace{-6pt}
\end{table*}

\subsection{Implementation Details}

We split each song into overlapping 10-second segments with a 5-second hop and aggregate segment-level predictions to obtain the song-level aesthetic score.

At inference time, after the HiGIA produces a score interval $[L_i,U_i]$ and an interpolation coefficient $\alpha_i$ for each segment $i$, we define a confidence weight as $w_i = 1/(U_i - L_i)$. Intuitively, a narrower interval reflects lower uncertainty and higher confidence derived from classifier consensus, allowing more reliable segments to contribute more to the final aggregation. The final aesthetic score is then obtained as weighted averages over segments.
\begin{equation}
    \label{eq:weighted_average}
    \hat{y}_{\text{song}} = \frac{\sum_{i} w_i[(1-\alpha_i)L_i + \alpha_iU_i]}{\sum_{i} w_i}.
\end{equation}

The data is split into train, validation, and test sets with an 8:1:1 ratio. Experiments are conducted on the original rating scales (5-point for SongEval and 100-point for the internal dataset). All audio is processed at 24 kHz. We use MDX-Net \cite{kim2021kuielab} as the source separation model and 3-layer Transformer encoders. For multi-dimensional evaluation, all aesthetic dimensions share the same feature extraction backbone but utilize independent HiGIA heads. For HiGIA, the number of score bins is set to $(K_{g_1}, K_{g_2}, K_{g_3}) = (2, 4, 8)$ for SongEval dataset and $(3, 5, 9)$ for internal dataset, to accommodate their distinct rating scales and annotation standards. We train with a batch size of 32 using Adam (learning rate $5 \times 10^{-4}$, $\epsilon = 10^{-8}, \beta_1=0.9,\beta_2=0.999$).



\subsection{Evaluation Metrics}

To quantify model performance, we use four metrics widely adopted in prior work \cite{huang2024voicemos, yao2025songeval, liu2025musiceval} to measure the prediction accuracy and rank consistency between the model's predictions and the human-annotated ground truth scores:

\begin{itemize}[leftmargin=*, itemsep=0pt, parsep=0pt, topsep=4pt]
    \item \textbf{Mean Squared Error (MSE)}: Measures the average squared difference between the predicted and ground truth scores. A lower value indicates better accuracy.
    \item \textbf{Linear Correlation Coefficient (LCC)}: Measures the degree of linear correlation between the predictions and ground truth. A higher value indicates better trend alignment.
    \item \textbf{Spearman Rank Correlation Coefficient (SRCC)}: Evaluates the consistency of the ordinal rank between predictions and ground truth. A higher value indicates better ranking ability.
    \item \textbf{Kendall’s Tau Rank Correlation (KTAU)}: Also measures rank correlation but is more robust in handling tied ranks. A higher value indicates stronger ranking ability.
\end{itemize}

\begin{table*}[!t]
  \centering
  \caption{Ablation study on SongEval and the Internal dataset. \textit{w/o MSAF} and \textit{w/o HiGIA} denote variants without the corresponding module. $\uparrow$/$\downarrow$ indicate higher/lower is better. \textbf{Bold} denotes the best result.}
  \label{tab:ablation-study}
  
  \renewcommand{\arraystretch}{0.88} 
  \small
  \setlength{\tabcolsep}{1.8pt} 
  
  \begin{tabular*}{\textwidth}{@{\extracolsep{\fill}} lccccc | lccccc }
    \toprule
    \multicolumn{6}{c}{\textbf{SongEval Dataset (5-point scale)}} & \multicolumn{6}{c}{\textbf{Internal Dataset (100-point scale)}} \\
    \cmidrule(r){1-6} \cmidrule(l){7-12}
    \textbf{Dim} & \textbf{Model} & \textbf{MSE$\downarrow$} & \textbf{LCC$\uparrow$} & \textbf{SRCC$\uparrow$} & \textbf{KTAU$\uparrow$} & 
    \textbf{Dim} & \textbf{Model} & \textbf{MSE$\downarrow$} & \textbf{LCC$\uparrow$} & \textbf{SRCC$\uparrow$} & \textbf{KTAU$\uparrow$} \\
    \midrule

    \multirow{3}{*}{Musicality}
      & w/o MSAF  & \textbf{0.237} & \textbf{0.903} & 0.903 & 0.737 & 
      \multirow{3}{*}{Overall} 
      & w/o MSAF  & 30.7 & 0.868 & 0.880 & 0.695 \\
      & w/o HiGIA & 0.284 & 0.901 & 0.899 & 0.735 & 
      & w/o HiGIA & 40.5 & 0.882 & 0.889 & 0.702 \\
      & \textbf{Ours} & 0.239 & 0.902 & \textbf{0.906} & \textbf{0.744} & 
      & \textbf{Ours} & \textbf{21.7} & \textbf{0.912} & \textbf{0.921} & \textbf{0.753} \\
    \midrule
    
    \multirow{3}{*}{Coherence}
      & w/o MSAF  & 0.289 & 0.876 & 0.881 & 0.714 & 
      \multirow{3}{*}{Singing} 
      & w/o MSAF  & 35.9 & 0.839 & 0.876 & 0.695 \\
      & w/o HiGIA & 0.297 & 0.883 & 0.887 & 0.720 & 
      & w/o HiGIA & 38.7 & 0.823 & 0.866 & 0.679 \\
      & \textbf{Ours} & \textbf{0.243} & \textbf{0.895} & \textbf{0.896} & \textbf{0.732} & 
      & \textbf{Ours} & \textbf{27.3} & \textbf{0.885} & \textbf{0.907} & \textbf{0.736} \\
    \midrule
    
    \multirow{3}{*}{Memorability}
      & w/o MSAF  & \textbf{0.317} & \textbf{0.883} & \textbf{0.878} & 0.702 & 
      \multirow{3}{*}{Melody} 
      & w/o MSAF  & \textbf{28.8} & \textbf{0.828} & 0.817 & 0.630 \\
      & w/o HiGIA & 0.322 & 0.882 & 0.875 & \textbf{0.705} & 
      & w/o HiGIA & 31.6 & 0.817 & 0.815 & 0.631 \\
      & \textbf{Ours} & 0.321 & 0.882 & 0.874 & 0.702 & 
      & \textbf{Ours} & 29.4 & \textbf{0.828} & \textbf{0.830} & \textbf{0.643} \\
    \midrule
    
    \multirow{3}{*}{Clarity}
      & w/o MSAF  & 0.304 & 0.876 & 0.872 & 0.692 & 
      \multirow{3}{*}{Arrangement} 
      & w/o MSAF  & 22.0 & 0.817 & \textbf{0.825} & 0.641 \\
      & w/o HiGIA & 0.303 & 0.886 & 0.881 & 0.703 & 
      & w/o HiGIA & \textbf{18.4} & \textbf{0.851} & \textbf{0.825} & 0.641 \\
      & \textbf{Ours} & \textbf{0.264} & \textbf{0.893} & \textbf{0.891} & \textbf{0.717} & 
      & \textbf{Ours} & 18.9 & 0.843 & 0.824 & \textbf{0.650} \\
    \midrule
    
    \multirow{3}{*}{Naturalness}
      & w/o MSAF  & \textbf{0.255} & \textbf{0.893} & \textbf{0.891} & \textbf{0.713} & 
      \multirow{3}{*}{Audio Quality} 
      & w/o MSAF  & 24.0 & 0.881 & 0.885 & 0.709 \\
      & w/o HiGIA & 0.257 & 0.891 & 0.886 & 0.712 & 
      & w/o HiGIA & 22.3 & 0.897 & 0.891 & 0.718 \\
      & \textbf{Ours} & 0.263 & 0.889 & 0.883 & 0.709 & 
      & \textbf{Ours} & \textbf{21.1} & \textbf{0.915} & \textbf{0.910} & \textbf{0.742} \\
    \midrule
    
    \multirow{3}{*}{Average}
      & w/o MSAF  & 0.280 & 0.886 & 0.885 & 0.712 & 
      \multirow{3}{*}{Average} 
      & w/o MSAF  & 28.3 & 0.847 & 0.857 & 0.674 \\
      & w/o HiGIA & 0.293 & 0.889 & 0.886 & 0.715 & 
      & w/o HiGIA & 30.3 & 0.854 & 0.857 & 0.674 \\
      & \textbf{Ours} & \textbf{0.266} & \textbf{0.892} & \textbf{0.890} & \textbf{0.721} & 
      & \textbf{Ours} & \textbf{23.7} & \textbf{0.877} & \textbf{0.878} & \textbf{0.705} \\
      
    \bottomrule
  \end{tabular*}
  \vspace{-6pt}
\end{table*}

\subsection{Experiment Results}

We conducted comparative experiments and ablation studies on the test sets of the two datasets. The results are analyzed as follows.

\subsubsection{Comparison Experiments}
We compared our model with the two baselines on the two datasets to evaluate its performance.




As shown in Table~\ref{tab:comparison-results}, our framework achieves consistently strong performance on the SongEval dataset and outperforms the baselines on most dimensions and metrics. In particular, it obtains the best results on Musicality, Coherence, and Clarity across all four metrics, demonstrating a clear advantage in modeling core musical attributes and overall perceptual consistency. These gains are especially evident in rank-based metrics, where our framework achieves the highest SRCC and KTAU on Musicality and Clarity, suggesting that it better captures relative human preference in dimensions that involve stronger subjective judgment. On Memorability, our framework remains competitive and achieves the best LCC while matching UTMOS on MSE and KTAU, indicating comparable performance on this dimension. On Naturalness, our model is slightly weaker than UTMOS across the four metrics. This is reasonable, since UTMOS is specifically designed for speech-related quality assessment and is therefore more sensitive to vocal cues such as breathing, phrasing, and voice naturalness.

On the internal dataset, our framework shows consistent advantages on the dimensions of Overall, Singing, and Melody, where it achieves the best performance across all four metrics. Notably, the gains on Overall and Singing are substantial, particularly in SRCC and KTAU, indicating that our method not only improves score prediction accuracy but also better preserves the relative ranking preferred by human annotators. For Audio Quality, although our framework does not achieve the lowest MSE, it obtains the highest LCC, SRCC, and KTAU, which suggests that it is more effective at capturing the perceptual trend and relative ranking of listeners. For Arrangement, our framework is slightly weaker than UTMOS in MSE, LCC, and SRCC, but achieves the best KTAU. 
%
%
This result indicates that under the high subjective uncertainty inherent in arrangement-related judgments, our method still maintains better ranking consistency with human preference.

Overall, our framework achieves the best average performance across all four metrics on both datasets, which cover both AI-generated and human-created songs. Compared with the strongest baseline, it improves the average KTAU from 0.713 to 0.721 on SongEval and from 0.671 to 0.705 on the internal dataset, while also obtaining the best average MSE, LCC and SRCC on both datasets. 
The gains are more evident on the internal dataset of human-created songs, suggesting that our method remains effective under more fine-grained evaluation settings. These results demonstrate that our framework provides robust and reliable song aesthetics evaluation across different data distributions and rating scales.


\subsubsection{Ablation Studies}
We conducted ablation studies by individually removing MSAF and HiGIA from our model, which are denoted as \textit{w/o MSAF} and \textit{w/o HiGIA} in the results, respectively.

As shown in Table~\ref{tab:ablation-study}, ablating either component reduces the overall average performance on both datasets, while the full model achieves the best average results across all four metrics. This indicates that both modules contribute positively to the complete framework, even when their effects vary across different aesthetic dimensions.

The effect of MSAF is particularly evident on dimensions sensitive to stem interaction, such as Coherence on SongEval and Overall on the internal dataset, where the full model yields clear gains in both correlation and ranking metrics. In contrast, for relatively isolated attributes such as Memorability, Naturalness, and Melody, removing MSAF sometimes leads to comparable or even slightly better results on individual metrics. This also suggests that the improvement brought by MSAF is not merely due to increased model capacity; rather, it is effective specifically because it models multi-stem musical interactions that are more relevant to certain aesthetic dimensions.

HiGIA also contributes consistently to the full model, especially on dimensions with stronger subjective uncertainty. Compared with the variant without HiGIA, the complete framework generally achieves better performance across all four metrics on highly subjective dimensions such as Musicality and Overall. These gains indicate that HiGIA effectively models the subjective uncertainty inherent in aesthetic judgments, leading to predictions that are more consistent with human preference rankings.

Taken together, these results show that MSAF and HiGIA play complementary roles: MSAF enhances the modeling of complex musical feature representations, while HiGIA improves robustness to subjective uncertainty in aesthetic scoring. Although the full model does not achieve the best result on every individual dimension, it consistently obtains the best average performance across all four metrics on both datasets. This demonstrates that the two components jointly contribute to improving overall song aesthetics evaluation, and validates the effectiveness of the proposed design.

\section{CONCLUSION}

We propose a song aesthetics evaluation framework that overcomes limitations of prior work. It learns complex musical features by fusing information across the mixture, vocal stem, and accompaniment stem via a Multi-Stem Attention Fusion module, and then utilizes Hierarchical Granularity-Aware Interval Aggregation that mirrors experts' coarse-to-fine decisions to achieve more stable and accurate predictions. Experiments on two diverse datasets covering AI-generated and human-created full-length songs show superior performance across multiple aesthetic dimensions compared with strong SOTA baselines. Ablation studies further validate the essential contributions of both modules in capturing complex musical interactions and modeling subjective uncertainty.





\bibliography{ISMIRtemplate}

%
%
%
%

\end{document}